\documentclass[a4paper]{jpconf}
\usepackage{graphicx}
\usepackage{color}
\usepackage{soul}

\newcommand{\text}[1]{\mathrm{#1}}

\newcommand{\al}{\alpha}

\newcommand{\eps}{\varepsilon}

\newcommand{\sig}{\sigma}

\newcommand{\tobs}{t_{\rm obs}}

\newcommand{\Tmc}{T_{\rm MC}}
\newcommand{\Ts}{T_{\rm s}}
\newcommand{\nlfs}{n_{\rm LFS}}

\begin{document}

\title{Devitrification of the Kob-Andersen glass former: Competition with the locally favored structure}
% in biased trajectories

\author{Francesco Turci}
\address{HH Wills Physics Laboratory, Tyndall Avenue, Bristol BS8 1TL, UK}
\address{Centre for Nanoscience and Quantum Information, Tyndall Avenue, Bristol BS8 1FD, UK}
\author{C. Patrick Royall}
\address{HH Wills Physics Laboratory, Tyndall Avenue, Bristol BS8 1TL, UK}
\address{Centre for Nanoscience and Quantum Information, Tyndall Avenue, Bristol BS8 1FD, UK}
\address{School of Chemistry, University of Bristol, Cantock's Close, Bristol BS8 1TS, UK}
\author{Thomas Speck}
\address{Institut f\"ur Physik, Johannes Gutenberg-Universit\"at Mainz, Staudingerweg 7-9, 55128 Mainz, Germany}

\ead{thomas.speck@uni-mainz.de}

\begin{abstract}
  Supercooled liquids are kinetically trapped materials in which the transition to a thermodynamically more stable state with long-range order is strongly suppressed. To assess the glass-forming abilities of a liquid empirical rules exist, but a comprehensive microscopic picture of devitrification is still missing. Here we study the crystallization of a popular model glass former, the binary Kob-Andersen mixture, in small systems. We perform trajectory sampling employing the population of the locally favored structure as order parameter. While for large population a dynamical phase transition has been reported, here we show that biasing towards a small population of locally favored structures induces crystallization, and we estimate the free energy difference. This result sheds new light on the competition between local and global structure in glass-forming liquids and its implications for crystallization.
\end{abstract}

%% ---- introduction ----

\section{Introduction}

Devitrification, the formation of ordered domains in a glass, is an issue of technological importance (\emph{e.g.} applications as diverse as optical fibers, DVDs, long-term nuclear waste storage) but also for model glass formers that have been devised to remain in a disordered liquid state even after being supercooled far below their melting temperature. These liquids are characterized by a dramatic increase of their structural relaxation time without a pronounced change of their two-point structure \cite{royall2015physrep}. Supercooled liquids thus challenge the view that the structure of a material underlies its static and dynamic properties, with competing theoretical explanations such as the energy landscape picture~\cite{adam1965,debenedetti2001}, random first-order theory~\cite{lubchenko2007}, and dynamic facilitation~\cite{chandler2010}. While strictly metastable, the lifetime of the liquid state in these systems is very large compared to the time window accessible to experiments and simulations. However, with the increase of computational power, in simulations the observable time span has expanded and, indeed, a number of model glass formers have been shown to spontaneously crystallize given enough time~\cite{Pedersen2007,toxvaerd2009}. These simulations allow valuable insights into the transformation kinetics and pathways, for example whether devitrification occurs through the nucleation of ordered domains~\cite{kelton1991} or ``avalanche'' dynamics~\cite{sanz2014}.

The Kob-Andersen (KA) binary mixture of large and small particles has been long considered a canonical model for simple supercooled liquids~\cite{kob1995}. Its robustness with respect to crystallization stems from the strong attractive interactions between small and large particles, leading to a frustration of the thermodynamic ground state. At zero temperature, the ground state has been argued to be a phase-separated state of a face centered cubic (FCC) crystal of the large particles in coexistence with an equimolar CsCl crystal~\cite{fernandez2003pre}. Obtaining numerically the high-density phase diagram of binary Lennard-Jones mixtures has been a long-standing challenge~\cite{lamm2001,toxvaerd2009,nandi2016}, and recent work \cite{pedersen2018} has shown that small variations of the composition lead to a rich behavior with several competing crystalline structures.

The KA liquid is a fragile glass-former that is characterized, in the low temperature regime, by a gradual change of the local structure~\cite{coslovich2007,malins2013fara}. This is measured in terms of number of particles forming specific geometric motifs that minimize the potential energy locally, also known as \textit{locally favored structures} (LFS)~\cite{royall2015physrep}. The fact that these structures do not lead to an ordered phase is rationalized through \emph{geometric frustration}: while locally favored structures minimize the free energy they cannot form large large domains due to frustration, in the sense that they either do not tile space or cannot be formed due to compositional constraints~\cite{tarjus2005,crowther2015}. In the case of the Kob-Andersen mixture, the LFS is the bicapped square antiprism~\cite{malins2013fara} which tiles space but is compositionally frustrated \cite{crowther2015}. Its population is negligible at high temperatures and increases to roughly 15\% as the liquid is cooled from the onset of slow-dynamics to the avoided mode coupling transition temperature $\Tmc$.

This model has been studied to show the emergence--in the supercooled regime--of a dynamical phase transition between inactive and active trajectories~\cite{hedges2009,speck2012jcp}. This transition has been shown to coincide with a transition between trajectories exceptionally rich/poor in LFS~\cite{speck2012,turci2017prx}. At low temperatures, the probability to explore LFS-rich trajectories is enhanced and the dynamical coexistence gap between LFS-rich and LFS-poor trajectories is reduced. This has motivated the proposal of possible alternative scenarios accounting for the low temperature fate of the dynamical phase transition~\cite{turci2017prx}, including the existence of a low-temperature critical point, similarly to what has been observed in softened kinetically constrained models~\cite{elmatad2010,elmatad2013space}. Here we discuss the interplay of the emerging crystallization with the trajectory sampling and the observed dynamical transition.

% Despite its relative robustness to crystallization, at temperatures hardly accessible to conventional simulations the mixture of large and small particles is prone to phase separate and crystallize through the formation of extended face centered cubic (FCC) domains of the large species, as reported in very long simulations and optimized calculations~\cite{toxvaerd2009}. 

\nocite{fullerton2014}

%% ---- model ----

\section{Model and methods}

\begin{figure}[t]
  \centering
  \includegraphics{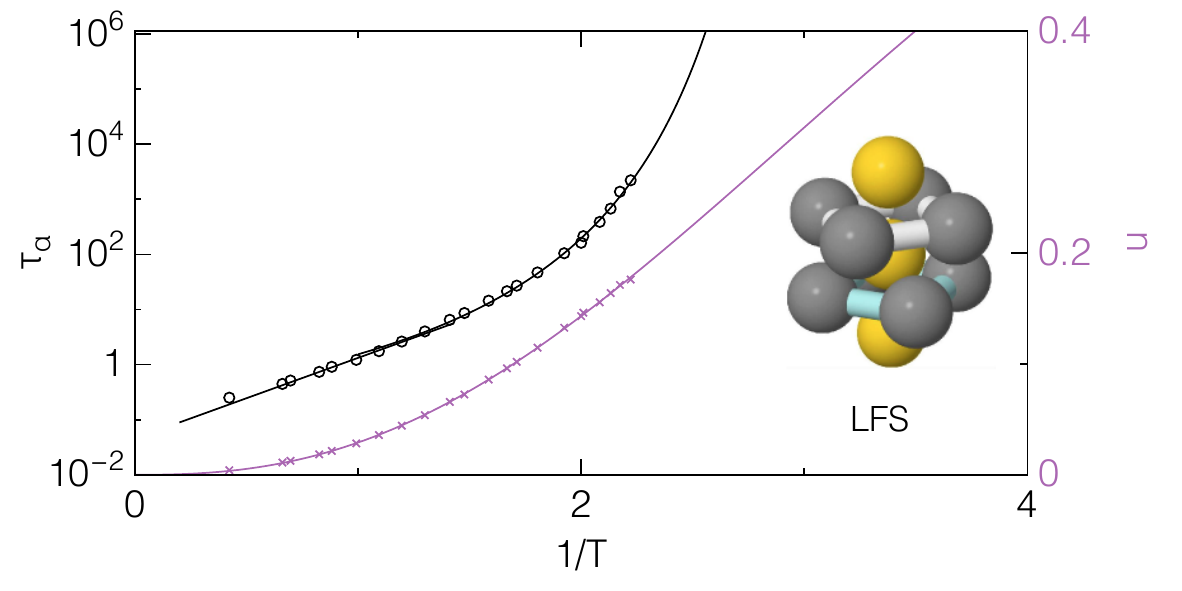}
  \caption{The KA model glass former. The structural relaxation time $\tau_\al$ (black symbols) strongly increases as the temperature is lowered. The solid black line is a fit to the Vogel-Fulcher-Tamman expression $\ln\tau_\al\propto(T-T_0)^{-1}$ which would diverge at $T_0\simeq0.3$. Concurrently with the increase of relaxation time, the population of a certain local motif (snapshot) absent in the normal liquid strongly increases (purple symbols). The purple line is a fit to the empirical Fermi function $n(T)=[1+(T/T_{1/2})^\al]^{-1}$ with $T_{1/2}\simeq0.25$ and exponent $\al\simeq2.5$.}
  \label{fig:KA}
\end{figure}

We consider the canonical Kob-Andersen binary mixture of large (A) and small (B) particles, which interact through the truncated and shifted Lennard-Jones potentials $u_{\al\beta}(r) =u_{\al\beta}^\text{LJ}(r)-u_{\al\beta}^\text{LJ}(2.5\sig_{\al\beta})$ for $r\leq 2.5\sig_{\al\beta}$ and zero for $r>2.5\sig_{\al\beta}$ with $u_{\al\beta}^\text{LJ}(r) = 4\eps_{\al\beta}[(\sig_{\al\beta}/r)^{12}-(\sig_{\al\beta}/r)^6]$. We employ the original parameters for length and energy: $\sig_\text{AA}=\sig$, $\sig_\text{AB}=0.8\sig$, $\sig_\text{BB}=0.88\sig$, $\eps_\text{AA}=\eps$, $\eps_\text{AB}=1.5\eps$, and $\eps_\text{BB}=0.5\eps$. Throughout the paper we employ reduced Lennard-Jones units with respect to the large particles, \emph{i.e.}, we measure length in units of $\sig$, energy in units of $\eps$, time in units of $\sqrt{m\sig^2/\eps}$ with particle mass $m$, and we set Boltzmann's constant to unity. Simulations are performed with stoichiometry 4A:1B in the isochoric-isothermal ensemble (NVT) at the conventional number density $\rho=N/V=1.2$. We generate initial trajectories via molecular dynamics simulations thermalized at temperature $\Ts$ via an Andersen thermostat.

% improve
Trajectory sampling is performed according to a Monte-Carlo scheme with moves inspired by transition path sampling techniques~\cite{bolhuis2002,speck2012jcp}. Discrete trajectories $x$ consist of $K=L+1$ configurations sampled at uniform time interval $\Delta t$. The total observation time $\tobs=L\Delta t$ is typically much larger than the structural relaxation time $\tau_{\alpha}$. In order to guarantee convergence and gather the required statistics, this approach is restricted to rather small system sizes, here we use $N=216$ particles. The sampling of trajectories is coupled to a well-tested parallel replica-exchange scheme: a biasing pseudo-potential in trajectory space drives the sampling of trajectories with exceptionally large/small values of the time-integrated total population of LFS,
\begin{equation}
  \mathcal{N}[x]=N \sum_{i=-L/2}^{L/2}\hat\nlfs(i).
\end{equation}
Here, $\hat\nlfs(i)$ is the fraction of particles participating in an LFS in configuration $i$. The pseudo-potential $\Psi$ used in the replica-exchange step is chosen to have a parabolic form around prescribed reference values $\{\mathcal{N}_0^j\}$
\begin{equation}
  \Psi_j(\mathcal{N}) =\frac{1}{2}\omega (\mathcal{N}-\mathcal{N}_0^j)^2,
\end{equation}
where the strength $\omega$ is chosen to ensure good mixing between replicas.  Note that the purpose of the biasing potential is to sample rare regions of trajectory space: the actual dynamics of the individual trajectories is not biased, and simply follows a traditional scheme to perform the molecular dynamics of a thermalized liquid. More details on the algorithm and its implementation can be found in Refs.~\cite{speck2012jcp,speck2012,turci2017prx}.

%% ---- results ----

\section{Results}

\subsection{Suppressing locally favored structures induces crystallization}

\begin{figure}[t]
  \centering
  \includegraphics[scale=1]{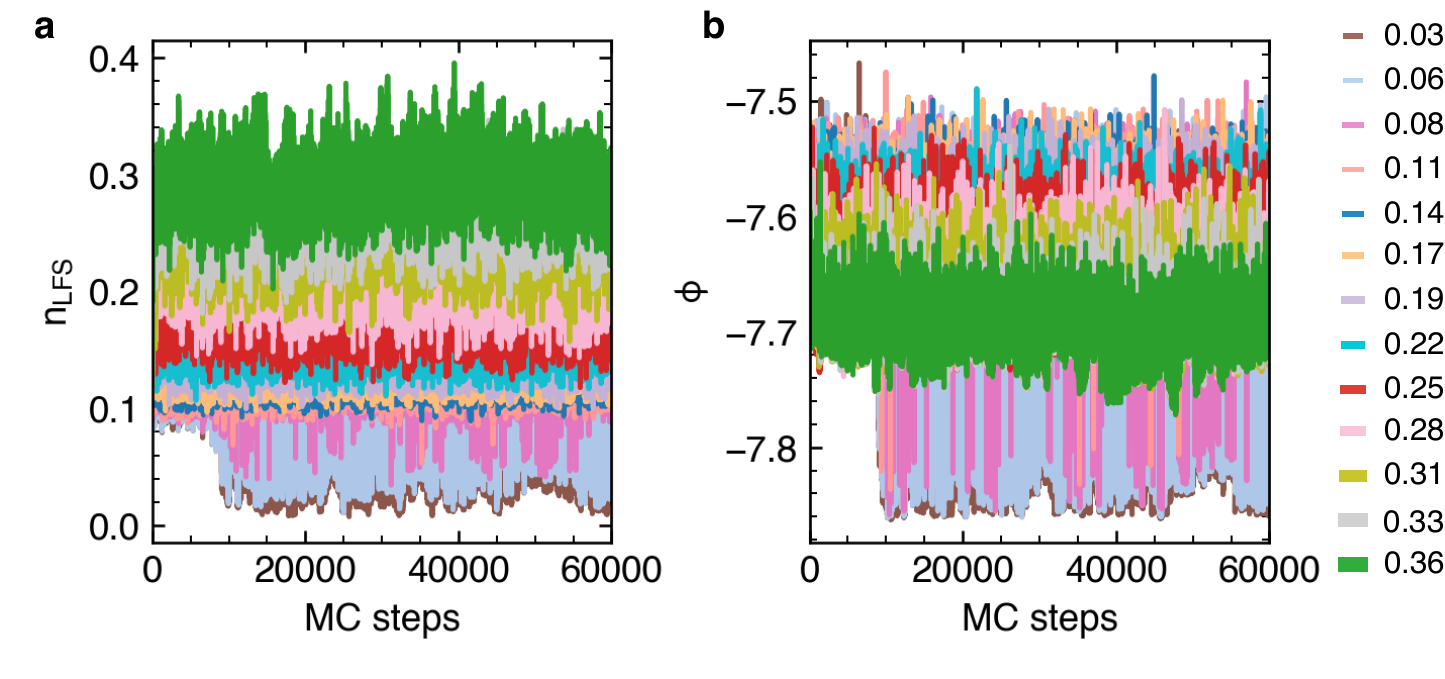}
  \caption{Replica-exchange Monte-Carlo in trajectory space. The different colors correspond to the different replicas with reference fraction $\nlfs^j=\mathcal{N}_0^j/NK$ (see legend) of particles bound in LFS. (a) Time-integrated fraction $\nlfs$ of LFS per trajectory. (b) Inherent state energy of the central configuration of each trajectory.}
  \label{fig:MonteCarlo}
\end{figure}

We now consider $R=13$ distinct replicas of trajectories of length $L=60$, which correspond to an observation time $\tobs\approx 9 \tau_{\alpha}$. Replicas are equally spaced with positions $\{\mathcal{N}_0^j \}=\{ 360, 720,\dots 4680 \}=NK\times\{0.028, 0.56,\dots,0.36\}$. Sampling of trajectory space across the $R$ replicas is preformed during $6\times10^4$ Monte Carlo sweeps, where 2500 swap attempts between all the replicas are performed at each step. The thermostat is set to a relatively low temperature $T=0.50$, which is above the avoided mode-coupling transition temperature $\Tmc=0.435$ but well below the temperature $T=0.60$ employed in previous studies~\cite{hedges2009,speck2012jcp,speck2012}.

\begin{figure}[t]
  \centering
  \includegraphics{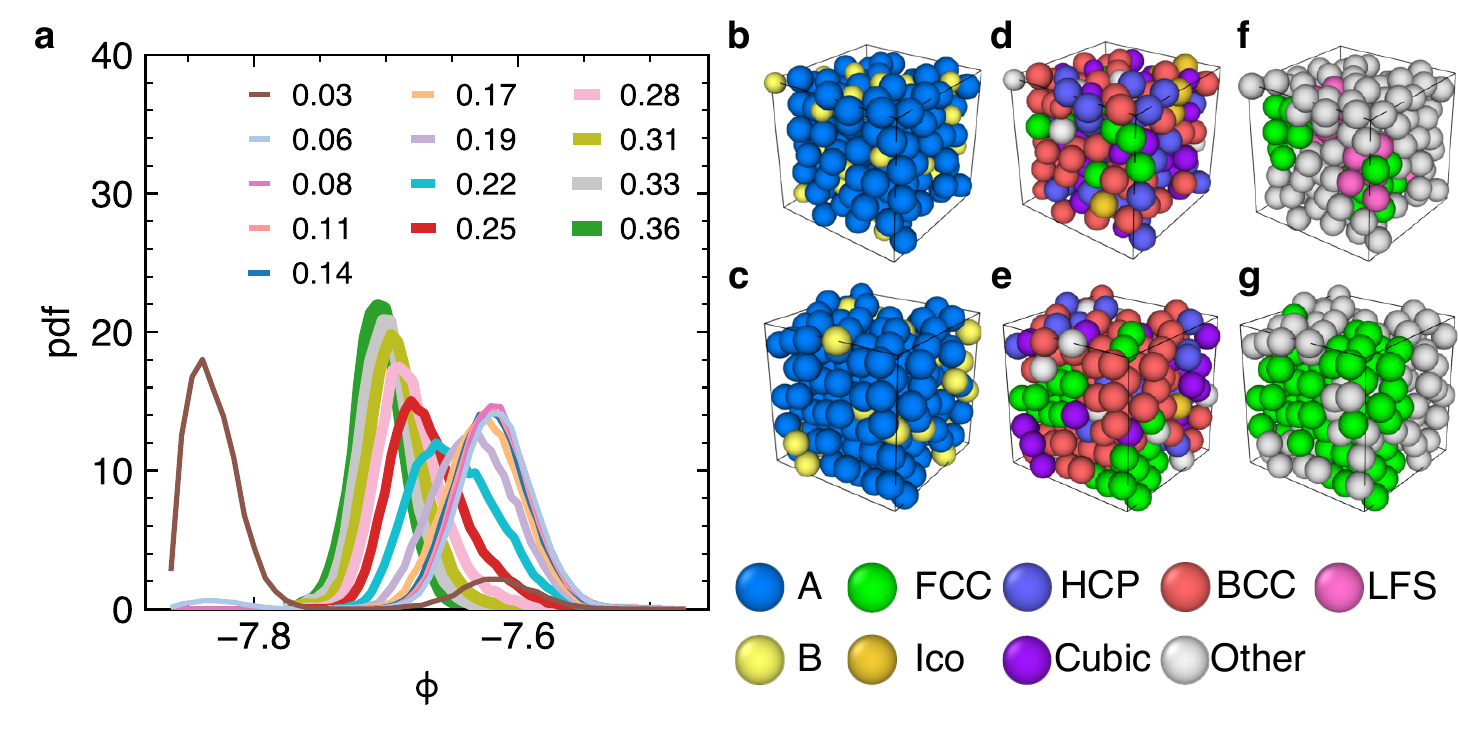}
  \caption{(a) Probability distributions of inherent state energies for the different replicas, labelled by the respective reference LFS fraction $\nlfs^j$. (b-c) Three-dimensional rendering of the large (A, blue) and small (B, yellow) particles in the liquid (b) and crystalline (c) states. (d-e)  As in (b-c) but color-coded according to the motifs detected by the Polyhedral Template Matching as implemented in the visualization software Ovito \cite{stukowski2009visualization}. (f-g) FCC and LFS regions as detected by the Topological Cluster Classification in the liquid (f) and crystal (g).}
  \label{fig:Hist}
\end{figure}

In Fig.~\ref{fig:MonteCarlo}, we illustrate the evolution of the Monte Carlo sampling in trajectory space for two representative observables: the time-integrated fraction of particles in LFS per trajectory $\nlfs=\mathcal{N}/(NK)$ and the potential energy $\phi$ of local minima (inherent states) per particle of the central configuration of every trajectory obtained via the \textsc{FIRE} minimization algorithm~\cite{bitzek2006}.
% cite Stillinger
We observe that after approximately 10.000 Monte-Carlo sweeps the replicas biased to fluctuate around the smallest concentrations of LFS show sudden large oscillations between very low populations ($\nlfs\approx 2\%$) and significantly larger values ($\nlfs\approx 10\%$). This corresponds to a dramatic change in the nature of the inherent states explored by the trajectories, as demonstrated by the probability distribution of the inherent states in Fig.~\ref{fig:Hist}(a). While replicas with reference fraction $\nlfs^j\geq0.11$ have inherent state energies well above $\phi=-7.8$, replicas with smaller reference values present long tails to much more negative inherent state energies. In particular, $\nlfs^j=0.03$ and $0.06$ display bimodal distributions with well separated high and low energy states at $\phi_{\rm high}\simeq-7.61$ and $\phi_{\rm low}\simeq-7.84$, respectively.

\subsection{Structure of the low energy states}

A structural analysis of the low energy states demonstrates that they correspond to a partially crystallized formation. We perform two independent analyses: on the one hand, we broadly discriminate between local crystalline order and icosahedral order employing the Polyhedral Template Matching (PTM) method~\cite{larsen2016robust}; on the other hand, we perform a more detailed analysis detecting the occurrence of crystalline motifs as well as local minima of the specific Kob-Andersen interaction potential, as compiled in the Topological Cluster Classification (TCC)~\cite{malins2013tcc,tccdownload}. 

The PTM method computes deviations between the local environment of a given atom with respect to template motifs (which are mainly crystalline) assigning the particles to the local structural motif with the smallest deviation. With no cutoff on the largest acceptable deviation, the method tends to detect false positives (for example overestimating the presence of crystalline structures). However, we are interested in relative differences between the crystalline and the liquid-like state, and since false-positives are equally detected in the two states, this will not affect our analysis.

In Fig.~\ref{fig:Hist}(b,c) we show that while globally a liquid-like and a crystal-like configuration show similar extended regions of large A particles spanning through the simulation box, the PTM analysis in Fig.~\ref{fig:Hist}(d,e) reveals nontrivial structural differences between the two states. While the overall populations of particles detected in hexagonal close packed (HCP), body centered cubic (BCC) and simple cubic regions is approximately unchanged (with typically large values due to false-positives of $\sim 50\%, \sim15\%$ and $\sim15\%$ respectively in both the liquid and crystal-like state) major differences are observed between the populations of face centered cubic (FCC) and icosahedral regions (Ico) which go from $\sim8\%$ to $\sim17\%$ (FCC) and from  $10\%$ to $\sim 1\%$. The FCC region, in particular, involves only large particles, indicating the fractionation of the system. Therefore, the PTM analysis provides a first indication that while in the liquid regime (i.e. when the population of LFS is above $10\%$) there is an important population of five-fold symmetric motifs this is suppressed when the system crystallizes.  

% TCC
A more detailed analysis of the structural changes taking place during crystallization is provided by the Topological Cluster Classification~\cite{malins2013tcc}. This ring-counting algorithm focuses on local motifs that are relevant for the specific Kob-Andersen interaction as they represent local minima of the potential for $n$ particles. Here we consider $9\leq n\leq 13$ and focus on a variety of structures. For instance, Fig.~\ref{fig:Hist}(f,g) shows that the TCC detects crystalline regions which mostly overlap with the PTM detection. However, the algorithm is also designed to detect the locally favored structure of the Kob-Anndersen mixture, which is present in the liquid-like configurations such as Fig.~\ref{fig:Hist}(f) but it is suppressed in the crystal-like ones, Fig.~\ref{fig:Hist}(g). More broadly, we employ the TCC to distinguish between structures that can be found in thermalized crystalline configurations of a Lennard-Jones system from structures that present some-degree of five-fold symmetry, i.e. composed of pentagonal rings.

\begin{figure}[t]
  \centering
  \includegraphics[width=\textwidth]{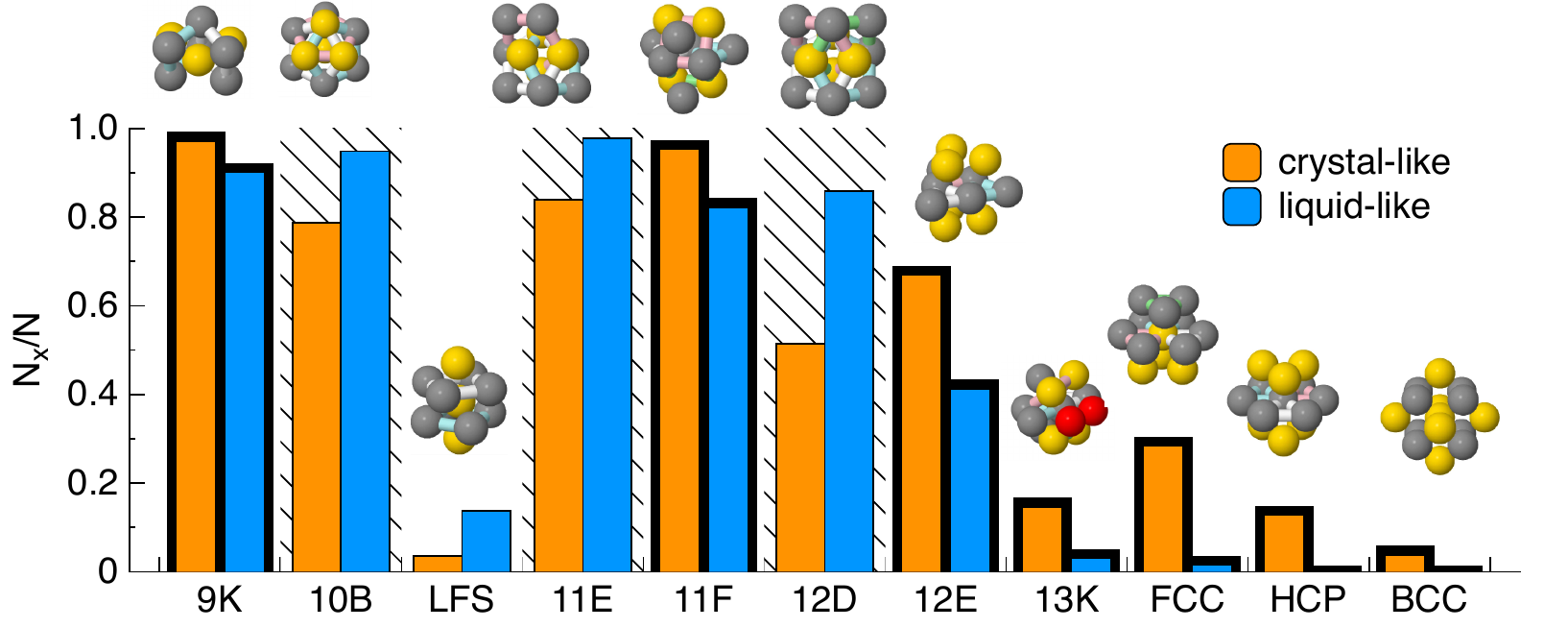}
  \caption{Structural spectrum from the Topological Cluster Classification (TCC) method for liquid-like and crystal-like trajectories (see main text for definition), representing the population of motif $x$ as the fraction of particles in motif $x$. Crystalline motifs are highlighted by thick lines, while five-fold symmetric structures have a dashed background. The number in the label corresponds to the number of particles in the motif, while the distinguishes between specific minima of the Kob-Andersen mixture (K) and minima for other potentials (such as the Morse potential). The locally favored structure of the Kob-Andersen mixture (the bicapped square antiprism) is denoted LFS.}
  \label{fig:TCC}
\end{figure}

In Fig.~\ref{fig:TCC}, we show the structural spectrum, as provided by the TCC, of the liquid-like and the crystal-like trajectories as extracted from the two corresponding representative replicas with $n_0^{\rm crystal}\simeq0.028$ and $n_0^{\rm liquid}\simeq0.14$. This shows that, as the system crystallizes, the population of all the five-fold symmetric structures (dashed background) decreases rapidly, with largest five-fold symmetric motif (12D) undergoing the most dramatic change. The same occurs, evidently, for the population of LFS. At the same time, all the crystalline descriptors increase, with the fraction of particles in FCC regions increasing of one order of magnitude. Hence, both the TCC and the PTM independently show that crystallization of the supercooled liquid occurs at the expenses of five-fold symmetric local order, similar to what has been reported for hard spheres~\cite{taffs2016}. This is, in fact, the expected behavior of the transition from a \textit{liquid} phase (where five-fold symmetry dominates) to a \textit{crystalline} phase. Specific to the Kob-Andersen model is that such a transition can be more easily accessed if the population of long-lived bicapped square antiprisms (which is the LFS and is not five-fold symmetric) is suppressed.

\subsection{Conventional sampling and barrier to nucleation}

\begin{figure}[t]
  \centering
  \includegraphics{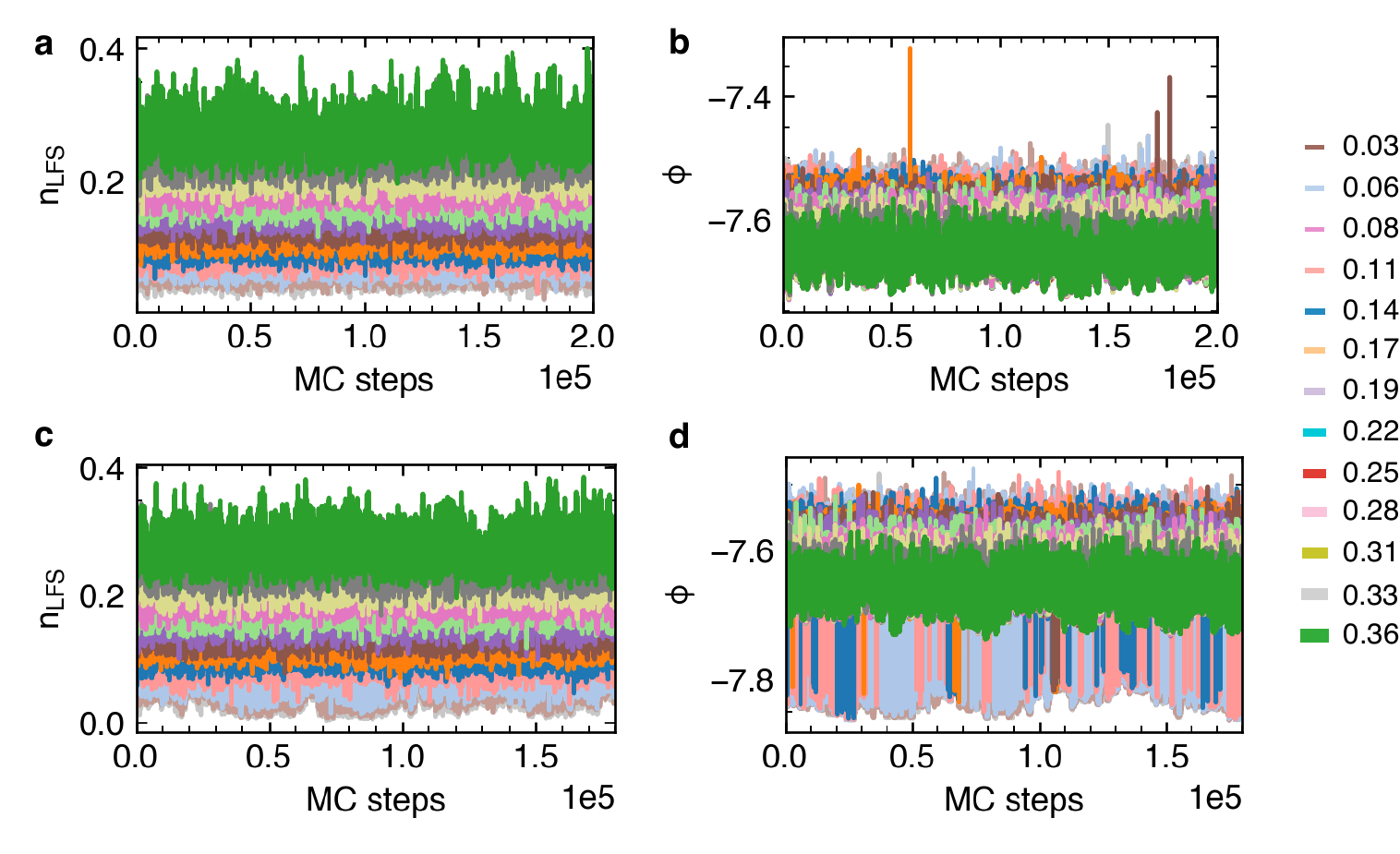}
  \caption{Conventional umbrella sampling: (a,c) population of LFS and (b,c) inherent state energies. (a,b) Show the Monte-Carlo sampling initialized with equilibrated configurations at $T=0.50$, while (c,d) show the results when the inital states are extracted from trajectories that have also crystallized during trajectory sampling. Colors correspond to reference fraction of LFS, see legend.}
  \label{fig:StandardMc}
\end{figure}

In the limit of very short trajectories, the replica exchange algorithm becomes equivalent to standard umbrella sampling along a static structural observables (the population of LFS), that we call \textit{conventional sampling}. Performing the sampling with trajectories of duration $\tobs=0.1\tau_{\alpha}$ at the same temperature $\Ts=0.50$, we find that the system is again robust against crystallization, and $2\cdot 10^5$ Monte-Carlo sweeps do not show signs of crystallization, with inherent state energies always above $\phi=-7.75$, see Fig.~\ref{fig:StandardMc}(a,b). 

In order to estimate the free energy difference between the crystal-like state (dominated by large particles arranged in FCC domains) and the supercooled liquid at $T=0.50$ we employ a different route. We select configurations from trajectories obtained via trajectory sampling of long trajectories (i.e. including the crystal-like formations discussed in the previous section) and from those configurations we subsequently restart the conventional sampling. This allows us to obtain extensive sampling of the transitions from the crystal-like to the liquid state, as shown in Fig.~\ref{fig:StandardMc}(c,d), stationary for about $1.8\cdot 10^5$ Monte Carlo steps.

\begin{figure}[t]
  \centering
  \includegraphics{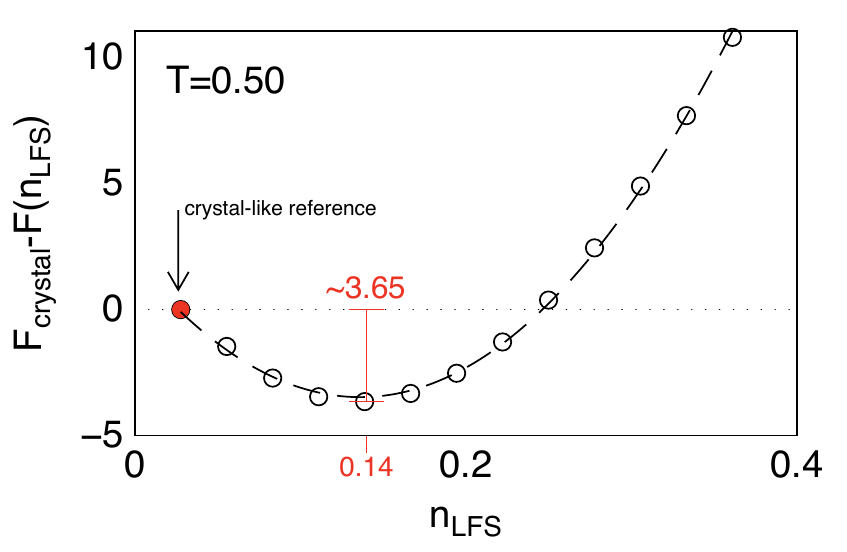}
  \caption{Free energy difference between the crystal-like state at $\nlfs=0.028$ and the amorphous states with larger populations of LFS. The minimum corresponds to the equilibrium population $\nlfs=0.14$ at temperature $T=0.50$. The dashed line is a parabolic fit.}
  \label{fig:freeEnergies}
\end{figure}

Through the Multistate Bennet Acceptance Ratio (MBAR) method~\cite{shirts2008statistically} we are able to compute the (relative) free energies associated to each individual replica and evaluate free energy differences along the biasing coordinate, i.e. the population of LFS. The result of this reweighting technique is shown in Fig.~\ref{fig:freeEnergies}. As expected, the free energy has a minimum at $\nlfs=0.14$, the equilibrium population of LFS for the considered sampling temperature $T=0.50$. The free energy difference (in reduced units) between the crystal-like state and the equilibrium liquid is $\Delta F=F_{\rm crystal}-F(\nlfs)\simeq-3.65$, more than four times smaller than the free energy difference between the equilibrium liquid and the amorphous phase with the largest LFS populations $\nlfs=0.36$. This relatively small difference is compatible with the spontaneous crystallization at lower temperatures $T=0.40$ reported in previous studies~\cite{toxvaerd2009} and recent very long simulations on Graphic Processing Units~\cite{ingebrigtsen2018}.

%% ---- summary ----

\section{Conclusions}

We have reported the crystallization of the KA model glass former. In contrast to previous work, crystallization is achieved through the suppression of the locally favored structure. However, we need to sample trajectories of sufficient length (larger than the structural relaxation time), whereas crystallization is again absent in conventional umbrella sampling. This supports the idea that not only the presence of the LFS but the dynamical correlations it induces are responsible for the presence and absence of effective kinetic constraints. As shown here, their absence allows the supercooled liquid to explore a larger manifold of relaxation paths, which eventually leads to the thermodynamically stable state, the crystal.

\ack
 CPR acknowledges the Royal Society for funding. FT and CPR acknowledge the European Research Council (ERC consolidator grant NANOPRS, project number 617266). This work was carried out using the computational facilities of the Advanced Computing Research Centre, University of Bristol.
%% ---- references ----

\section*{References}
\providecommand{\newblock}{}


\begin{thebibliography}{10}
\expandafter\ifx\csname url\endcsname\relax
  \def\url#1{{\tt #1}}\fi
\expandafter\ifx\csname urlprefix\endcsname\relax\def\urlprefix{URL }\fi
\providecommand{\eprint}[2][]{\url{#2}}

\bibitem{royall2015physrep}
Royall C~P and Williams S~R 2015 {\em Phys. Rep.\/} {\bf 560} 1

\bibitem{adam1965}
Adam G and Gibbs J 1965 {\em J. Chem. Phys.\/} {\bf 43} 139--146

\bibitem{debenedetti2001}
Debenedetti P and Stillinger F 2001 {\em Nature\/} {\bf 410} 259--67

\bibitem{lubchenko2007}
Lubchenko V and Wolynes P 2007 {\em Annu. Rev. Phys. Chem.\/} {\bf 58} 235--266

\bibitem{chandler2010}
Chandler D and Garrahan J~P 2010 {\em Annu. Rev. Condens. Matter Phys.\/} {\bf
  61} 191--217 ISSN 1545-1593

\bibitem{Pedersen2007}
Pedersen U~R, Bailey N~P, Dyre J~C and Schr{\o}der T~B 2007 {\em
  arXiv:0706.0813\/}

\bibitem{toxvaerd2009}
Toxvaerd S, Pedersen U~R, Schroder T~B and Dyre J~C 2009 {\em J. Chem. Phys.\/}
  {\bf 130} 224501

\bibitem{kelton1991}
Kelton K~F 1991 {\em Solid State Phys.\/} {\bf 45} 75--177

\bibitem{sanz2014}
Sanz E, Valeriani C, Zaccarelli E, Poon W~K, Cates M and Pusey P~N 2014 {\em
  Proc. Nat. Acad. Sci.\/} {\bf 111} 75--80

\bibitem{kob1995}
Kob W and Andersen H 1995 {\em Phys. Rev. E\/} {\bf 51} 4626--4641

\bibitem{fernandez2003pre}
Fern\'{a}ndez J~R and Harrowell P 2003 {\em Phys. Rev. E\/} {\bf 67} 011403

\bibitem{lamm2001}
Lamm M~H and Hall C~K 2001 {\em Fluid Phase Equilibria\/} {\bf 182} 37--46

\bibitem{nandi2016}
Nandi U~K, Banerjee A, Chakrabarty B and Bhattacharyya S~M 2016 {\em J. Chem.
  Phys.\/}

\bibitem{pedersen2018}
Pedersen U~R, Schr{\o}der T~B and Dyre J~C 2018 {\em arXiv:1803.08956\/}

\bibitem{coslovich2007}
Coslovich D and Pastore G 2007 {\em J. Chem. Phys\/} {\bf 127} 124504

\bibitem{malins2013fara}
Malins A, Eggers J, Tanaka H and Royall C~P 2013 {\em Faraday Discussions\/}
  {\bf 167} 405--423

\bibitem{tarjus2005}
Tarjus G, Kivelson S~A, Nussinov Z and Viot P 2005 {\em J. Phys.: Condens.
  Matter\/} {\bf 17} R1143--R1182

\bibitem{crowther2015}
Crowther P, Turci F and P R~C 2015 {\em J. Chem. Phys.\/} {\bf 143} 044503

\bibitem{hedges2009}
Hedges L~O, Jack R~L, Garrahan J~P and Chandler D 2009 {\em Science\/} {\bf
  323} 1309--1313

\bibitem{speck2012jcp}
Speck T and Chandler D 2012 {\em J. Chem. Phys.\/} {\bf 136} 184509

\bibitem{speck2012}
Speck T, Malins A and Royall C~P 2012 {\em Phys. Rev. Lett.\/} {\bf 109} 195703

\bibitem{turci2017prx}
Turci F, Royall C~P and Speck T 2017 {\em Phys. Rev. X\/} {\bf 7} 031028

\bibitem{elmatad2010}
Elmatad Y~S, Jack R~L, Chandler D and Garrahan J~P 2010 {\em Proc. Nat. Acad.
  Sci.\/} {\bf 107} 12793--12798

\bibitem{elmatad2013space}
Elmatad Y~S and Jack R~L 2013 {\em J. Chem. Phys.\/} {\bf 138} 12A531

\bibitem{fullerton2014}
Fullerton C~J and Jack R~L 2014 {\em Phys. Rev. Lett.\/} {\bf 112} 255701

\bibitem{bolhuis2002}
Bolhuis P~G, Chandler D, Dellago C and Geissler P~L 2002 {\em Ann. Rev. Phys.
  Chem.\/} {\bf 53} 291--318

\bibitem{bitzek2006}
Bitzek E, Koskinen P, G{\"a}hler F, Moseler M and Gumbsch P 2006 {\em Phys.
  Rev. Lett.\/} {\bf 97} 170201

\bibitem{stukowski2009visualization}
Stukowski A 2009 {\em Modelling and Simulation in Materials Science and
  Engineering\/} {\bf 18} 015012

\bibitem{larsen2016robust}
Larsen P~M, Schmidt S and Schi{\o}tz J 2016 {\em Modelling and Simulation in
  Materials Science and Engineering\/} {\bf 24} 055007

\bibitem{malins2013tcc}
Malins A, Williams S~R, Eggers J and Royall C~P 2013 {\em J. Chem. Phys.\/}
  {\bf 139} 234506

\bibitem{tccdownload}
Malins A, Williams S, Turci F, Eggers J and Royall C~P 2015 Topological cluster
  classification download
  \urlprefix\url{http://www.chm.bris.ac.uk/pt/paddy/tcc.html}

\bibitem{taffs2016}
Taffs J and Royall C~P 2016 {\em Nat. Commun.\/} {\bf 7} 13225

\bibitem{shirts2008statistically}
Shirts M~R and Chodera J~D 2008 {\em J. Chem. Phys.\/} {\bf 129} 124105

\bibitem{ingebrigtsen2018}
Ingebrigtsen T~S, Dyre J~C, Schroder T~B and Royall C~P 2018 {\em preprint\/}

\end{thebibliography}
\end{document}